\documentclass[final,3p,times,sort&compress]{elsarticle}

\usepackage{amsmath}
\usepackage{amssymb}
\usepackage{epsfig}

\journal{Physics Letters B}

\begin{document}

\begin{frontmatter}



\title{The ratio $R={\rm d}\sigma_L/{\rm d}\sigma_T$ in heavy-quark pair leptoproduction \\
as a probe of linearly polarized gluons in unpolarized proton}


\author[dubna]{A.V.~Efremov}
\ead{efremov@theor.jinr.ru}
\author[yerphi,dubna2]{N.Ya.~Ivanov\corref{cor1}}
\cortext[cor1]{Corresponding author.}
\ead{nikiv@yerphi.am}
\author[dubna,dubna2]{O.V.~Teryaev}
\ead{teryaev@theor.jinr.ru}

\address[dubna]{Bogoliubov Laboratory of Theoretical Physics, JINR, 141980 Dubna, Russia}
\address[yerphi]{Yerevan Physics Institute, Alikhanian Br.~2, 0036 Yerevan, Armenia}
\address[dubna2]{Veksler and Baldin Laboratory of High Energy Physics, JINR, 141980 Dubna, Russia}

\begin{abstract}
We study the Callan-Gross ratio $R={\rm d}\sigma_L/{\rm d}\sigma_T$ in heavy-quark pair leptoproduction, $lN\rightarrow l^{\prime}Q\bar{Q}X$, as a probe of linearly polarized gluons inside unpolarized proton, where ${\rm d}\sigma_T$ (${\rm d}\sigma_L$) is the differential cross section of the $\gamma^*N\rightarrow Q\bar{Q}X$ process initiated by a transverse (longitudinal) virtual photon. Note first that the maximal value for the quantity $R$ allowed by  the photon-gluon fusion with unpolarized gluons is large, about 2. We calculate the contribution of the transverse-momentum dependent gluonic counterpart of the Boer-Mulders function, $h_{1}^{\perp g}$, describing the linear polarization of gluons inside unpolarized proton. Our analysis shows that the maximum value of the ratio $R$ depends strongly on the gluon polarization; it varies from 0 to $\frac{Q^2}{4m^2}$ depending on $h_{1}^{\perp g}$. We conclude that the Callan-Gross ratio in heavy-quark pair leptoproduction is predicted to be large and very sensitive to the contribution of linearly polarized gluons. For this reason, future measurements of the longitudinal and transverse components of the charm and bottom production cross sections at the proposed EIC and LHeC colliders seem to be very promising for determination of the linear polarization of gluons inside unpolarized proton.

\end{abstract}


\end{frontmatter}


\section{Introduction and notation} 
\label{1.0}
It is well-known that the heavy flavor production in lepton-proton DIS offers direct probe of the gluon distributions in the proton. In particular, the transverse momentum dependent (TMD) density of the linearly polarized gluons inside unpolarized proton, $h_{1}^{\perp g}(\zeta,\vec{k}_{T}^2)$, can directly be measured in heavy-quark pair electroproduction. This is because the density $h_{1}^{\perp g}(\zeta,\vec{k}_{T}^2)$ is $T$- and chiral-even (contrary to its quark analogue, so-called Boer-Mulders function $h_{1}^{\perp q}(\zeta,\vec{k}_{T}^2)$, which can only be detected in pairs with an other $T$-odd quantity).

The TMD distributions of polarized partons in unpolarized nucleon, $h_{1}^{\perp q}$ and $h_{1}^{\perp g}$, have been introduced in Refs.~\cite{Boer-Mulders} and \cite{Mulders_2001}, respectively.
In Refs.~\cite{Boer_HQ_1,Boer_HQ_2,Boer_HQ_3,we9,we_TMD}, azimuthal correlations in heavy-quark pair production in unpolarized electron-proton collisions as probes of the density $h_{1}^{\perp g}$ have been considered at leading order (LO) in QCD.\footnote{Concerning the opportunities to measure the function $h_{1}^{\perp g}$ in unpolarized hadron-hadron collisions, see review \cite{Boer_2015}. Some ideas and models concerning the origin/generation of the linearly polarized gluons in unpolarized proton can be found in Refs.~\cite{Fenya_2011,Catani_2012,Boer_2012,Metz_2011,Skokov_2017}.} In particular, it was shown that the azimuthal $\cos \varphi$ and $\cos 2\varphi$ asymmetries are predicted to be large and very sensitive to the function $h_{1}^{\perp g}$ \cite{we_TMD}. (Here $\varphi$ is the heavy quark (or anti-quark) azimuthal angle.)

In the present paper, we study the contribution of linearly polarized gluons to the Callan-Gross ratio $R={\rm d}\sigma_L/{\rm d}\sigma_T$ in the reaction $lN\rightarrow l^{\prime}Q\bar{Q}X$ with  unpolarized initial states, where ${\rm d}\sigma_L$ (${\rm d}\sigma_T$) is the differential cross section of the $\gamma^*N\rightarrow Q\bar{Q}X$ process initiated by a longitudinal (transverse) virtual photon. Our analysis shows that the quantity $R$ is expected to be large in wide kinematic ranges and very sensitive to the density $h_{1}^{\perp g}(\zeta,\vec{k}_{T}^2)$. We conclude that the Callan-Gross ratio in heavy-quark pair production in DIS could also be good probe of the linear polarization of gluons in unpolarized proton.

In Refs.\cite{Boer_HQ_1,Boer_HQ_2,Boer_HQ_3}, it was proposed to study the linearly polarized gluons in unpolarized nucleon using the heavy-quark pair production in the reaction
\begin{equation} \label{1}
l(\ell )+N(P)\rightarrow l^{\prime}(\ell -q)+Q(p_{Q})+\bar{Q}(p_{\bar{Q}})+X(p_{X}). 
\end{equation}
To describe this process, the following hadron-level variables are used:
\begin{align}
\bar{S}&=2\left( \ell\cdot P\right), & y&=\frac{q\cdot P}{\ell\cdot P },& T_{1}&=\left(P-p_{Q}\right)^{2}-m^{2},& S&=\left(q+P\right)^{2}  \notag \\
Q^{2}&=-q^{2}, & x&=\frac{Q^{2}}{2q\cdot P},& U_{1}&=\left( q-p_{Q}\right)^{2}-m^{2}, & z&=-\frac{T_{1}}{2q\cdot P},  \label{2}
\end{align}
where $m$ is the heavy-quark mass.

To probe a TMD distribution, the momenta of both heavy quark and anti-quark, $\vec{p}_{Q}$ and $\vec{p}_{\bar{Q}}$, in the process (\ref{1}) should be measured (reconstructed). For further analysis, the sum and difference of the transverse heavy quark momenta are introduced,
\begin{align} \label{3}
\vec{K}_{\perp}&=\frac{1}{2}\left(\vec{p}_{Q\perp}-\vec{p}_{\bar{Q}\perp}\right), &
\vec{q}_{T}&=\vec{p}_{Q\perp}+\vec{p}_{\bar{Q}\perp},
\end{align}
in the plane orthogonal to the direction of the target and the exchanged photon. The azimuthal angles of $\vec{K}_{\perp}$ and $\vec{q}_{T}$ (relative to the the lepton scattering plane projection, $\phi_l=\phi_{l^{\prime}}=0$) are denoted by $\phi_{\perp}$ and $\phi_{T}$,  respectively. 

\begin{figure*}
\begin{center}
\mbox{\epsfig{file=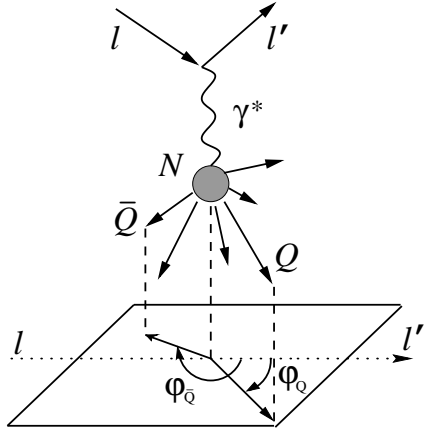,width=200pt}}
\caption{\label{Fg.1}\small Definition of the azimuthal angles $\varphi_Q$ and $\varphi_{\bar{Q}}$ in the nucleon rest frame.}
\end{center}
\end{figure*}

Following Refs.~\cite{Boer_HQ_1,Boer_HQ_2,Boer_HQ_3}, we use the approximation when $\vec{q}_{T}^2\ll \vec{K}_{\perp}^2$ and the outgoing heavy quark and anti-quark are almost back-to-back in the transverse plane, see Fig.~\ref{Fg.1}. In this case, the magnitudes of transverse momenta of the heavy quark and anti-quark are practically the same, $\vec{p}^2_{Q\perp}\simeq \vec{p}^2_{\bar{Q}\perp}\simeq\vec{K}_{\perp}^2$.  

At LO, ${\cal O}(\alpha _{em}\alpha_{s})$, the only parton-level subprocess for the reaction (\ref{1}) is the photon-gluon fusion:
\begin{equation} \label{5}
\gamma^{*}(q)+g(k_g)\rightarrow Q(p_{Q})+\bar{Q}(p_{\bar{Q}}), 
\end{equation}
where 
\begin{align} \label{6}
k_{g}^\mu &\simeq \zeta P^\mu + k_{T}^\mu,& \zeta=-\frac{U_1}{y\bar{S}+T_1} &=\frac{q\cdot k_g}{q\cdot P}.
\end{align}
The corresponding parton-level invariants are:
\begin{align} \label{7}
\hat{s}&=(q+k_g)^{2}\simeq\frac{m^2+\vec{K}_{\perp}^2}{z\,(1-z)},& t_{1}&=(k_g-p_{Q})^{2}-m^{2}\simeq\zeta T_1\simeq -z\,(\hat{s}+Q^2),&  u_{1}&=U_1\simeq -(1-z\,)(\hat{s}+Q^2).
\end{align}

\section{Production cross section} 
\label{2.0}

Schematically, the contribution of the photon-gluon fusion to the reaction (\ref{1}) has the following factorized form:
\begin{equation} \label{8}
{\rm d}\sigma\propto L(\ell,q)\otimes \Phi_g(\zeta, k_{T})\otimes \left| H_{\gamma^*g\rightarrow Q\bar{Q}X} (q,k_{g},p_{Q},p_{\bar{Q}})\right|^2, 
\end{equation}
where $L^{\alpha\beta}(\ell,q)=-Q^2 g^{\alpha\beta}+2(\ell^{\alpha}\ell^{\prime\beta}+\ell^{\beta}\ell^{\prime\alpha})$ is the leptonic tensor and $H_{\gamma^*g\rightarrow Q\bar{Q}X}(q,k_{g},p_{Q},p_{\bar{Q}})$ is the amplitude for the hard partonic subprocess. The convolutions $\otimes$ stand for the appropriate integration and traces over the color and Dirac indices.  

Information about parton densities in unpolarized nucleon is formally encoded in  corresponding TMD parton correlators. In particular, the gluon correlator is usually parameterized  as \cite{Mulders_2001}
\begin{equation} \label{9}
\Phi_g^{\mu\nu}(\zeta, k_{T})\propto - g_T^{\mu\nu}f_{1}^{g}\big(\zeta,\vec{k}_{T}^2\big)+\left(g_T^{\mu\nu}-2\frac{k_T^\mu k_T^\nu}{k_T^2}\right)\frac{\vec{k}_{T}^2}{2m^2_N}h_{1}^{\perp g}\big(\zeta,\vec{k}_{T}^2\big), 
\end{equation}   
where $m_N$ is the nucleon mass, 
\begin{align} \label{10}
g_T^{\mu\nu}&=g^{\mu\nu}- \frac{P^\mu n^\nu+P^\nu n^\mu}{P\cdot n},& n^\mu &=\frac{q^\mu+xP^\mu}{P\cdot q} .
\end{align}
In Eq.~(\ref{10}), the tensor $-g_T^{\mu\nu}$ is (up to a factor) the density matrix of unpolarized gluons. The TMD distribution $h_{1}^{\perp g}\big(\zeta,\vec{k}_{T}^2\big)$ describes  the contribution of linearly polarized gluons. The degree of their linear polarization is determined by the quantity $r=\frac{\vec{k}_{T}^2 h_{1}^{\perp g}}{2m^2_N f_{1}^{g}}$. In particular, the gluons are completely polarized along the $\vec{k}_{T}$ direction at $r=1$.\footnote{The TMD densities  under consideration have to satisfy the positivity bound \cite{Mulders_2001}: $\frac{\vec{k}_{T}^2}{2m^2_N}\big|h_{1}^{\perp g}(\zeta,\vec{k}_{T}^2)\big|\leq f_{1}(\zeta,\vec{k}_{T}^2)$.}

The LO predictions for the azimuth dependent cross section of the reaction (\ref{1}) are presented  in Ref.~\cite{Boer_HQ_2} as follows:
\begin{eqnarray} 
\frac{{\rm d}^{7}\sigma}{{\rm d}y\,{\rm d}x\,{\rm d}z\,{\rm d}^2\vec{K}_{\perp}{\rm d}^2\vec{q}_{T}}={\cal N}\Big\{A_0+A_1\cos\phi_{\perp}+A_2\cos 2\phi_{\perp}+\vec{q}_{T}^2\Big[B_0\cos 2(\phi_{\perp}-\phi_T)+B_1\cos (\phi_{\perp}-2\phi_T) \nonumber \\
+B_1^{\prime}\cos (3\phi_{\perp}-2\phi_T)+B_2\cos 2\phi_T+B_2^{\prime}\cos 2(2\phi_{\perp}-\phi_T)\Big]\Big\}, \label{11}
\end{eqnarray}
where ${\cal N}$ is a normalization factor, $\phi_{\perp}$ and $\phi_T$ denote the azimuthal angles of $\vec{K}_{\perp}$ and $\vec{q}_{T}$, respectively. 
The quantities $A_i$ ($i=0,1,2$) are determined by the unpolarized TMD gluon distribution, $A_i\sim f_{1}^{g}$, while $B_i$ ($i=0,1,2$) and $B_{1.2}^{\prime}$ depend on the linearly polarized gluon density, $B_i^{(\prime)}\sim h_{1}^{\perp g}$.

In the previous paper \cite{we_TMD}, we have recalculated the cross section for the reaction (\ref{1}) and our results for $A_i$, $B_i$ and $B_{1,2}^{(\prime)}$ do coincide with the corresponding ones presented in Ref.~\cite{Boer_HQ_2}.\footnote{The only exception is an evident misprint in Eq.(25) in Ref.\cite{Boer_HQ_2}. Note also a typo in Eq.(19): instead of ${\rm d}y_i=\frac{{\rm d}z_i}{z_1 z_2}$ should be ${\rm d}y_i=\frac{{\rm d}z_i}{z_i}$.} 

In the present letter, we are interested in the Callan-Gross ratio and have to integrate Eq.~(\ref{11}) over the quark and anti-quark azimuthal angles. Taking into account that 
\begin{align} \label{12}
{\rm d}^2\vec{K}_{\perp}{\rm d}^2\vec{q}_{T}&=\frac{X\,{\rm d}\vec{K}_{\perp}^2\,{\rm d}\vec{q}_{T}^2\,{\rm d}\varphi_Q\,{\rm d}\alpha}{4|\cos\alpha|\sqrt{\cos^2\alpha-X^2}},& X&=\frac{\vec{K}_{\perp}^2-\vec{q}_{T}^2/4}{\vec{K}_{\perp}^2+\vec{q}_{T}^2/4},& \alpha &=\pi-(\varphi_{Q}-\varphi_{\bar{Q}}), 
\end{align}
the production cross section has the following form at $\vec{q}_{T}^2\ll \vec{K}_{\perp}^2$:
\begin{align} 
\frac{{\rm d}^{5}\sigma^{(\pi)}}{{\rm d}y\,{\rm d}x\,{\rm d}z\,{\rm d}\vec{K}_{\perp}^2{\rm d}\vec{q}_{T}^2}&=\frac{\alpha_{em}}{4\pi}\frac{1}{xy}\frac{y^2}{1-\varepsilon}\Bigg\{\frac{{\rm d}^{3}\sigma_T}{{\rm d}z\,{\rm d}\vec{K}_{\perp}^2{\rm d}\vec{q}_{T}^2}+\varepsilon\frac{{\rm d}^{3}\sigma_L}{{\rm d}z\,{\rm d}\vec{K}_{\perp}^2{\rm d}\vec{q}_{T}^2}\Bigg\} \notag \\
&=\frac{\pi e_{Q}^{2}\alpha_{em}^2\alpha_{s}}{4(1-\varepsilon)\bar{S}^2}\frac{f_{1}^{g}(\zeta,\vec{q}_{T}^2)\hat{B}_T}{xy\, \zeta z\,(1-z)}\Bigg\{\left(1-2r \frac{\hat{B}^h_T}{\hat{B}_T}\right)+\varepsilon\frac{\hat{B}_L}{\hat{B}_T}\left(1-2r \frac{\hat{B}^h_L}{\hat{B}_L}\right)\Bigg\}.  \label{13}
\end{align}
Here $e_Q$ is the heavy quark charge, while ${\rm d}^3\sigma_T$ and ${\rm d}^3\sigma_L$ are the transverse and longitudinal components of the differential cross section of the $\gamma^*N\rightarrow Q\bar{Q}X$ process. The quantity $\varepsilon$ measures the degree of the longitudinal polarization of the virtual photon in the Breit frame \cite{Dombey},
\begin{align} \label{14}
\varepsilon &=\frac{2(1-y)}{1+(1-y)^2},& \zeta =\frac{-U_1}{y\bar{S}+T_1} &=x+ \frac{m^2+\vec{K}_{\perp}^2}{z\,(1-z)y\,\bar{S}},& r\equiv r(\zeta, \vec{q}_{T}^2)&=\frac{\vec{q}_{T}^2}{2m^2_N}\frac{h_{1}^{\perp g}\big(\zeta,\vec{q}_{T}^2\big)}{f_{1}\big(\zeta,\vec{q}_{T}^2\big)}.
\end{align}  

The superscript $^{(\pi)}$ on the differential cross section ${\rm d}^{5}\sigma^{(\pi)}$ means that we perform integration over $\varphi_{\bar{Q}}$ only in region of $|\,\varphi_{Q}-\varphi_{\bar{Q}}|\sim \pi$. The region of $|\,\varphi_{Q}-\varphi_{\bar{Q}}|\sim 0$ is out of the scope of this paper.

The coefficients $\hat{B}_i$ $(i=T,L)$ in Eq.~(\ref{13}) originate from the contribution of unpolarized gluons, while the quantities $\hat{B}^h_i$ are associated with the function $h_{1}^{\perp g}$. The LO predictions for $\hat{B}_i$ and $\hat{B}^h_i$ $(i=T,L)$ are given by\footnote{Note that the quantities $\hat B_T$ and $\hat B_L$ in Eq.~(\ref{15}) are related to $\hat B_2$ in Eq.~(16) of Ref.~[7] as follows: $\hat B_2=\hat B_T+\hat B_L$. Correspondingly, $\hat B_2^h=\hat B_T^h+\hat B_L^h$.}
\begin{align} 
\hat{B}_T\left(z,\vec{K}_{\perp}^2,Q^2\right)&=\frac{1-2l_z}{l_z}+\frac{4\,\hat{k}^2(l_z+\lambda-1/2)}{(\hat{k}^2+l_z+\lambda )^2},&  \hat{B}^h_T\left(z,\vec{K}_{\perp}^2,Q^2 \right)&=\frac{1}{2}\hat{B}_T\left(z,\vec{K}_{\perp}^2,Q^2\right)-\frac{1-2l_z}{2l_z},  \notag \\
\hat{B}_L\left(z,\vec{K}_{\perp}^2,Q^2 \right)&=\frac{8\,\hat{k}^2 l_z}{(\hat{k}^2+l_z+\lambda)^2},& \hat{B}^h_L\left(z,\vec{K}_{\perp}^2,Q^2 \right)&=\frac{1}{2}\hat{B}_L\left(z,\vec{K}_{\perp}^2,Q^2 \right),  \label{15} 
\end{align}
where the following notations are used:
\begin{align} \label{16}
l_z&=z\,(1-z),& \hat{k}^2&=\frac{\vec{K}_{\perp}^2}{Q^2},& \lambda &=\frac{m^2}{Q^2}.
\end{align} 

The quantities $A_0$ and $B_0$ in Eq.~(\ref{11}) are related to $\hat{B}_{T,L}$ and $\hat{B}^{h}_{T,L}$  defined by Eqs.~(\ref{15}) as follows:
\begin{align}
A_0&=\frac{y^2\kappa_f}{1-\varepsilon}\left(\hat{B}_T+\varepsilon\hat{B}_L \right),& \kappa_f&=\frac{m^2+\vec{K}_{\perp}^2}{2\xi y\,\bar{S}}e_{\!Q}^2 f_1^{g},  \notag \\
B_0&=\frac{y^2\kappa_h}{1-\varepsilon}\left(\hat{B}_T^h+\varepsilon\hat{B}_L^h \right),& \kappa_h&=\frac{m^2+\vec{K}_{\perp}^2}{2\xi y\,\bar{S}m_N^2}e_{\!Q}^2 h_1^{\perp g}. \label{17a} 
\end{align}

\section{The Callan-Gross ratio} 
\label{3.0}

One can see from Eq.(\ref{13}) that the gluonic version of the Boer-Mulders function, $h_{1}^{\perp g}(\zeta, \vec{q}_{T}^2)$, can, in principle, be determined from measurements of the $\vec{q}_{T}^2$-dependence of the Callan-Gross ratio, ${\rm d}^3\sigma_L/{\rm d}^3\sigma_T$. Let us first discuss the pQCD predictions for the ratio $R$ in the case when the unpolarized gluons only contribute to the reaction (\ref{1}), i.e. for $r=\frac{\vec{q}_{T}^2\, h_{1}^{\perp g}}{2 m^2_N\, f_{1}}=0$. At fixed values of $Q^2$, the corresponding quantity is:
\begin{equation} \label{17}
R\left(z,\vec{K}_{\perp}^2\right)=\frac{{\rm d}^3\sigma_L}{{\rm d}^3\sigma_T}\left(z,\vec{K}_{\perp}^2,r=0\right)= \frac{\hat{B}_L}{\hat{B}_T}\left(z,\vec{K}_{\perp}^2\right)=\frac{8\,\hat{k}^2 l_z^2}{\hat{k}^4 (1-2l_z)+2\lambda\, \hat{k}^2+(1-2l_z)(l_z+\lambda)^2},
\end{equation} 
where ${\rm d}^3\sigma_{L(T)}\equiv\frac{{\rm d}^{3}\sigma_{L(T)}}{{\rm d}z\,{\rm d}\vec{K}_{\perp}^2{\rm d}\vec{q}_{T}^2}$ and the variables $l_z$, $\hat{k}^2$ and $\lambda$ are defined by Eq.~(\ref{16}).

Our analysis shows that the function $R\,(z,\vec{K}_{\perp}^2)$ has an extremum  at $l_z=1/4$ and $\hat{k}^2=\lambda+1/4$ (i.e. at $z=1/2$ and $\vec{K}_{\perp}^2\!=m^2+Q^2/4$). This maximum value is:
\begin{equation} \label{18}
R\left(z=1/2,\vec{K}_{\perp}^2\!=m^2+Q^2/4\right)=\frac{2}{1+12\lambda}.
\end{equation}
Eq.~(\ref{18}) implies that the maximum value of the ratio $R$ at high $Q^2\gg m^2$ (i.e. for $\lambda\rightarrow 0$) is large, about 2. 

\begin{figure}
\begin{center}
\begin{tabular}{cc}
\mbox{\epsfig{file=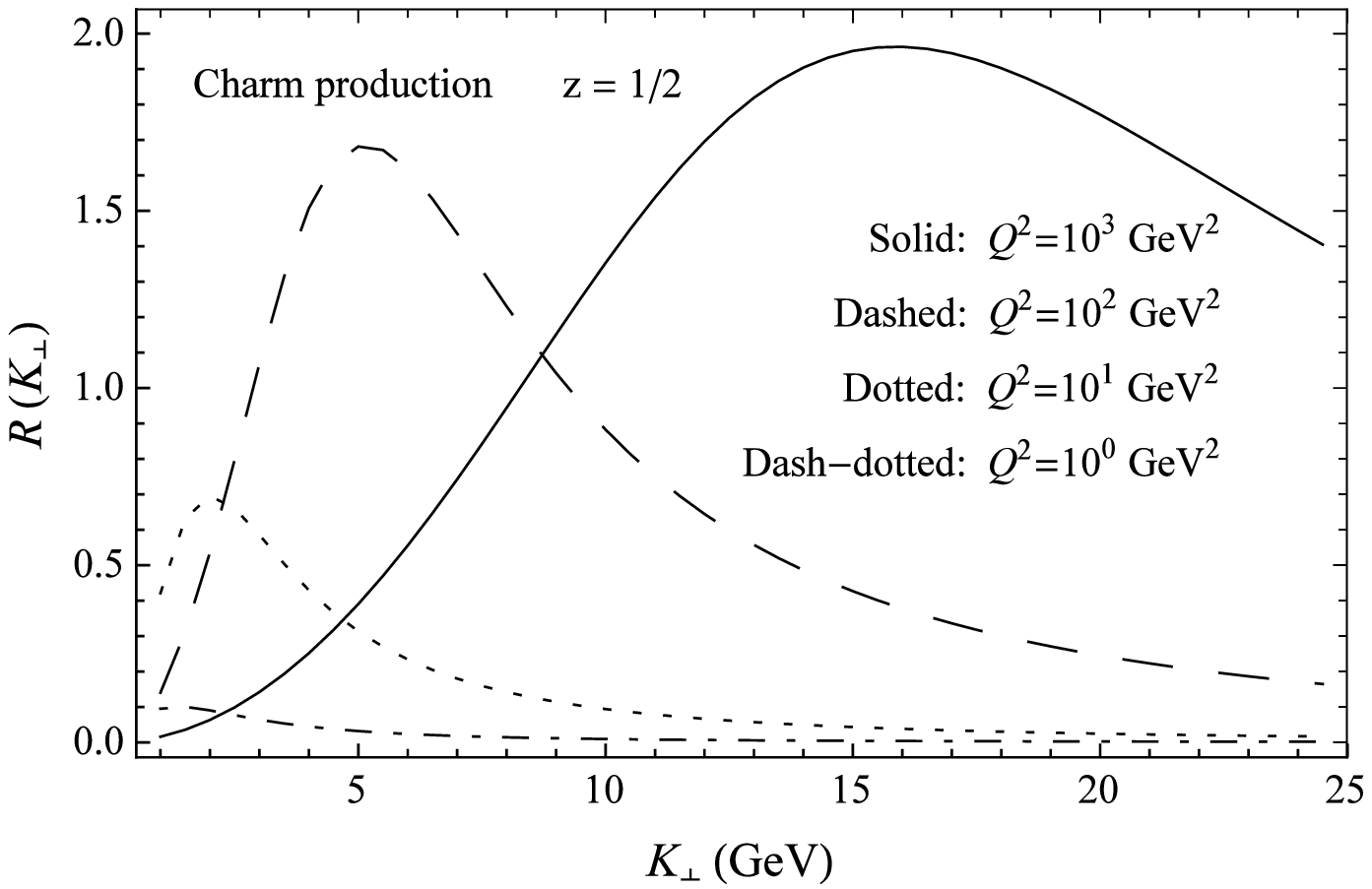,width=220pt}}
& \mbox{\epsfig{file=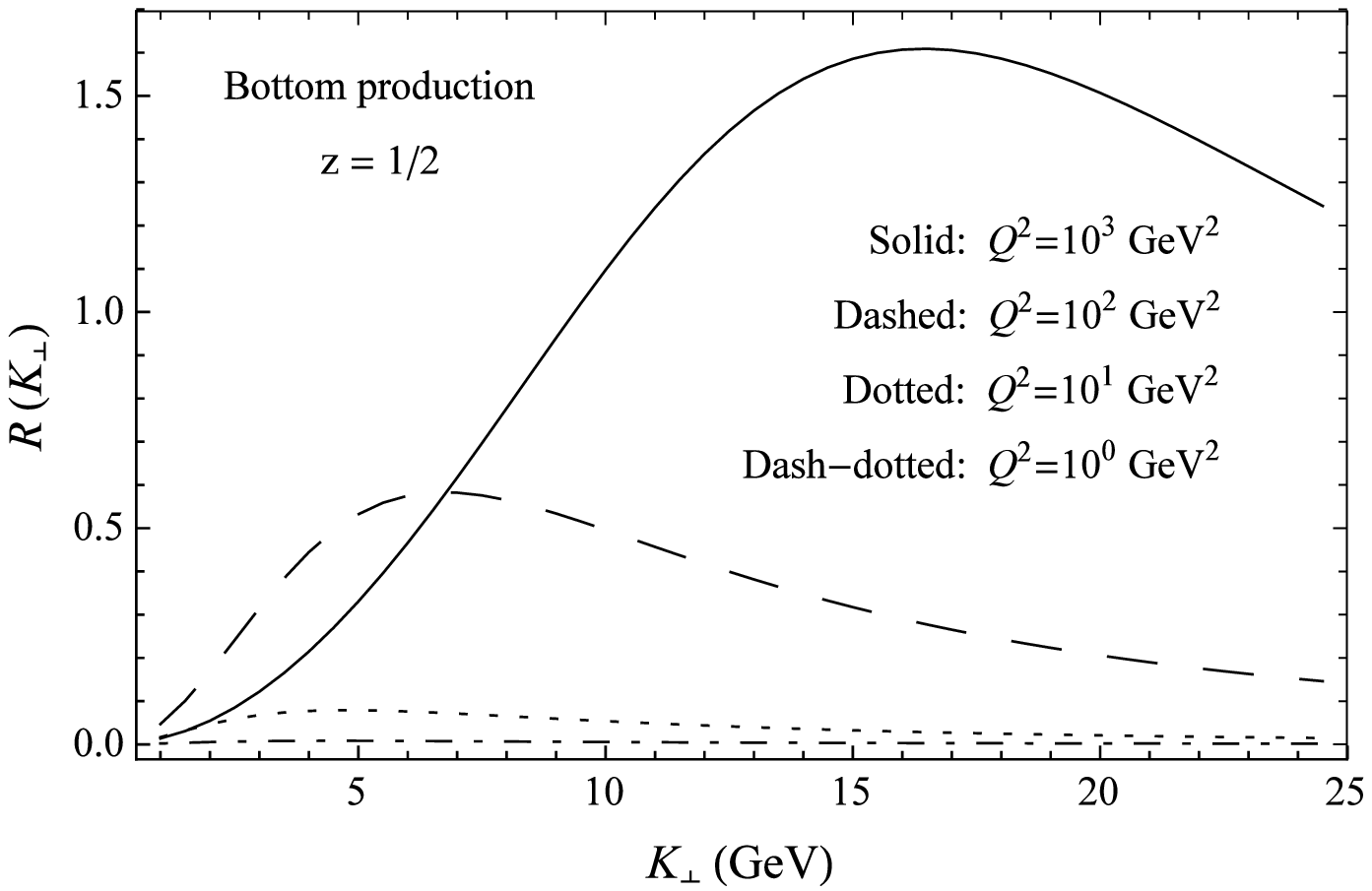,width=220pt}}\\
\end{tabular}
\caption{\label{Fg.2}\small Callan-Gross ratio $R\,(K_{\perp})\equiv R\,(z=1/2,K_{\perp})$ in charm ({\it left panel}) and bottom ({\it right panel}) electroproduction as a function of $K_{\perp}=\big|\vec{K}_{\perp}\big|$ at several values of $Q^2$.}
\end{center}
\end{figure}

The LO predictions for the quantity $R\,(K_{\perp})\equiv R\,(z=1/2,K_{\perp})$ in charm (left panel) and bottom (right panel) production as a function of $K_{\perp}=\big|\vec{K}_{\perp}\big|$ at several values of $Q^2$ are presented in Fig.~\ref{Fg.2}. In the present  paper, we use $m_c=$ 1.25 GeV and $m_b=$ 4.5 GeV. One can see from Fig.~\ref{Fg.2} that sizable values for the Callan-Gross ratio are expected in wide regions of $K_{\perp}$ and $Q^2$.

\begin{figure}[t]
\begin{center}
\mbox{\epsfig{file=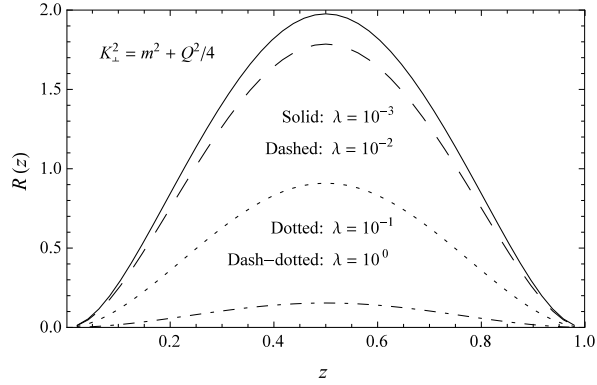,width=220pt}}
\caption{\label{Fg.3}\small Callan-Gross ratio $R\,(z)\equiv R\,(z,\vec{K}_{\perp}^2\!=m^2+Q^2/4)$ in heavy quark leptoproduction as a function of $z$ at several values of $\lambda$.}
\end{center}
\end{figure}

The quantity $R\,(z)\equiv R\,(z,\vec{K}_{\perp}^2\!=m^2+Q^2/4)$ depends only on $z$ and $\lambda$. Fig.~\ref{Fg.3} shows the Callan-Gross ratio $R\,(z)$ in heavy quark leptoproduction as a function of $z$ at several values of $\lambda$. One can see that the maximum value of $R\,(z)$ is accessed at $z=1/2$ and increases with $Q^2$.

Let us now discuss the contribution of the linearly polarized gluons to the Callan-Gross ratio. One can see from Eq.~(\ref{13}) that, at fixed values of $Q^2$, the corresponding quantity $R^h(z,\vec{K}_{\perp}^2,r)$, containing the contributions of both $f_1^g$ and $h_1^{\perp g}$ densities, is a function of three variables: 
\begin{equation} \label{19}
R^h\left(z,\vec{K}_{\perp}^2,r\right)=\frac{{\rm d}^3\sigma_L}{{\rm d}^3\sigma_T}\left(z,\vec{K}_{\perp}^2,r\right)=\frac{\hat{B}_L}{\hat{B}_T}\frac{1-2r\hat{B}^h_L\Big/\hat{B}_L}{1-2r\hat{B}^h_T\Big/\hat{B}_T}.
\end{equation}
Our analysis shows that the function $R^h(z,\vec{K}_{\perp}^2,r)$ has a maximum at $z=1/2$ and $\vec{K}_{\perp}^2=m^2+Q^2/4$ for all values of $r$ in the interval $-1\leq r\leq 1$. We find 
\begin{equation} \label{20}
R^h(r)\equiv R^h\left(z=1/2,\vec{K}_{\perp}^2=m^2+Q^2/4,r\right)=\frac{2(1-r)}{1+r+4\lambda\,(3-r)},
\end{equation}
where $r=\frac{\vec{q}_{T}^2}{2m^2_N}\frac{h_{1}^{\perp g}(\zeta_m,\vec{q}_{T}^2)}{f_{1}(\zeta_m,\vec{q}_{T}^2)}$ describes the degree of the linear polarization of gluons and $\zeta_m=2x\,(1+4\lambda)$.

\begin{figure}
\begin{center}
\mbox{\epsfig{file=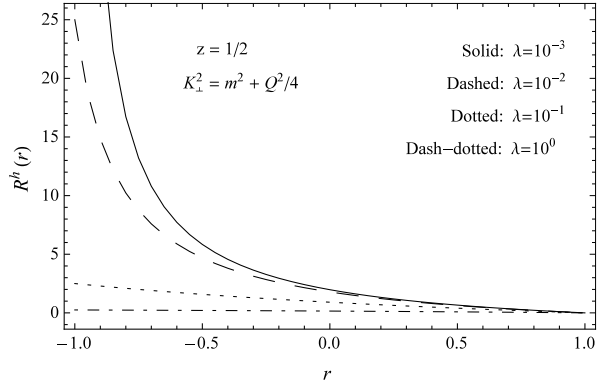,width=220pt}}
\caption{\label{Fg.4}\small Maximum value of the Callan-Gross ratio with the contribution of linearly polarized gluons, $R^h(r)$, as a function of $r$ at several values of $\lambda$.}
\end{center}
\end{figure}

The function $R^h(r)$ is depicted in Fig.~\ref{Fg.4} where its strong dependence on the variable $r$ is seen. In particular, $R^h(r)$ vanishes for $r=1$ because the longitudinal component of the cross section (\ref{13}), ${\rm d}^3\sigma_L$, is proportional to $(1-r)$. One can also see an unlimited growth of the Callan-Gross ratio with $Q^2$ at $r\rightarrow-1$, $R^h(r=-1)=\frac{Q^2}{4m^2}$. This is due to the fact that the transverse component of the cross section, ${\rm d}^3\sigma_T$, vanishes in the chiral limit for $z=1/2$, $\vec{K}_{\perp}^2=m^2+Q^2/4$ and $r=-1$.

In Fig.~\ref{Fg.4}, the Callan-Gross ratio is depicted at $z=1/2$, $\vec{K}_{\perp}^2=m^2+Q^2/4$ where it predicted to be maximal within pQCD. Note however that the gluon polarization can, in principle, be determined experimentally from measurements of the $\vec{q}_{T}^2$-dependence of the ratio ${\rm d}^3\sigma_L/{\rm d}^3\sigma_T$ in arbitrary kinematic. Our analysis shows that the quantity $R^h(z,\vec{K}_{\perp}^2,r)$ defined by Eq.~(\ref{19}) is an unambiguous function of $r$ for any (physical) values of $z$, $\vec{K}_{\perp}^2$ and $Q^2$. For this reason, comparing future experimental values of the Callan-Gross ratio, $R^{\,\rm exp}(z,\vec{K}_{\perp}^2,\vec{q}_{T}^2)$, with the predicted ones, $R^{\,\rm exp}(z,\vec{K}_{\perp}^2,\vec{q}_{T}^2)=R^h(z,\vec{K}_{\perp}^2,r)$, one could unambiguously estimate the quantity $r=\frac{\vec{q}_{T}^2\, h_{1}^{\perp g}}{2 m^2_N\, f_{1}}$ for given $z$, $\vec{K}_{\perp}^2$, $Q^2$ and $\vec{q}_{T}^2$:\footnote{Due to confinement, the heavy quark momenta can be determined/reconstructed from data with an accuracy not better than ${\cal O}\left(\Lambda_{\rm QCD}\right)$. For this reason, too small values of $|\vec{q}_{T}|\sim \Lambda_{\rm QCD}$ seem to be inaccessible.}
\begin{equation} \label{21}
r=\frac{1}{2}\frac{R^{\,\rm exp}\,\hat{B}_T-\hat{B}_L}{R^{\,\rm exp}\,\hat{B}_T^h-\hat{B}_L^h}=\frac{8\, l_z^2 Q^2 \vec{K}_{\perp}^2-R^{\,\rm exp}\left\{(1-2l_z)\left[ \vec{K}_{\perp}^4+(l_z Q^2+m^2)^2\right]+2m^2 \vec{K}_{\perp}^2\right\}}{2l_z \vec{K}_{\perp}^2\left\{4l_zQ ^2+R^{\,\rm exp}\left[(1-2l_z) Q^2-2m^2 \right] \right\}},
\end{equation}
with $l_z=z\,(1-z)$.

We conclude that the Callan-Gross ratio in heavy flavor leptoproduction is predicted to be large and very sensitive to the contribution of linearly polarized gluons in unpolarized proton. For this reason, this observable can be good probe of the gluonic counterpart of the Boer-Mulders function, $h_{1}^{\perp g}(\zeta,\vec{q}_{T}^2)$.

\section{Conclusions and Outlook} 
\label{4.0}

In this paper, we study the Callan-Gross ratio $R={\rm d}^3\sigma_L/{\rm d}^3\sigma_T$ in heavy-quark pair leptoproduction, $lN\rightarrow l^{\prime}Q\bar{Q}X$, as a probe of the linearly polarized gluon density, $h_{1}^{\perp g}(\zeta,\vec{k}_{T}^2)$, inside unpolarized proton. Here ${\rm d}^3\sigma_{L(T)}\equiv\frac{{\rm d}^{3}\sigma_{L(T)}}{{\rm d}z\,{\rm d}\vec{K}_{\perp}^2{\rm d}\vec{q}_{T}^2}$ is the  longitudinal (transverse) component of the differential cross section of the $\gamma^*N\rightarrow Q\bar{Q}X$ process.
We found that the maximal value of the ratio $R$ allowed by  the photon-gluon fusion with unpolarized gluons (i.e. when $h_{1}^{\perp g}$=\,0) is large, about 2, at LO in QCD. Our analysis shows that the quantity $R$ is  very sensitive to the contribution of $h_{1}^{\perp g}(\zeta,\vec{k}_{T}^2)$. In particular, the upper bound on the Callan-Gross ratio in presence of the linearly polarized gluons is $\frac{Q^2}{4m^2}$, i.e., it increases indefinitely  with $Q^2$. 

Our results could have very useful consequence: large experimental values of the Callan-Gross ratio, $R>2$, will directly indicate not only the presence of linearly polarized gluons but also the negative sign of $h_{1}^{\perp g}$. Note that negative values of the density $h_{1}^{\perp g}(\zeta,\vec{k}_{T}^2)$ correspond to the case when the gluon polarization direction is preferably orthogonal to $\vec{k}_{T}$.

Physically, our main observation can be formulated as follows. When a linearly polarized gluon interacts with transverse virtual photon, the heavy-quark production plane is preferably  orthogonal to the direction of the gluon polarization. On the contrary, the longitudinal component of the cross section $g^{\uparrow}\gamma^{\,*}\rightarrow Q\bar{Q}$ takes its maximum value when the  momenta of emitted quarks and the gluon polarization lie in the same plane. Moreover, the longitudinal cross section vanishes in the case when the gluon polarization is orthogonal to the heavy-quark production plane. 

We conclude that the ratio $R={\rm d}^3\sigma_L/{\rm d}^3\sigma_T$ in heavy-quark pair  leptoproduction could be good probe of the linear polarization of gluons inside unpolarized nucleon. 

Our analysis of the ratio $R$ is performed at LO in QCD. Unfortunately, radiative corrections to the process $\gamma^{\,*}g^{\uparrow}\rightarrow Q\bar{Q}$ (i.e. with linearly polarized initial gluons) are presently unavailable. 

The exact next-to-leading order (NLO) corrections to heavy flavor electroproduction with unpolarized initial states (quarks and gluons) have been computed in Ref.~\cite{Neerven_HQ_93}. 
As was noted in \cite{we7,we8}, the Callan-Gross ratio in heavy-quark leptoproduction is sufficiently stable under these corrections. In particular, large NLO contributions to the cross sections $\sigma_L$ and $\sigma_T$ cancel each other in their ratio $\frac{\sigma_L}{\sigma_T}(x,Q^2)$ with 10\% accuracy in the energy range $x>10^{-4}$. For this reason, it is of special interest to calculate the exact NLO corrections to $Q\bar{Q}$ leptoproduction for the case of linearly polarized initial gluons within the TMD factorization scheme. Perhaps, these (potentially large) contributions will also cancel each other in the ratio $R$.

As to the higher order contributions, they are usually estimated with the help of so-called soft-gluon resummation at energies not so far from the production threshold. Soft-gluon (or threshold) resummation for heavy flavor production in DIS with unpolarized and longitudinally polarized initial states has been performed in Refs.~\cite{Laenen_res_99} and \cite{Moch_res_00}, respectively. Resummations of soft-gluon logarithms for the azimuth-dependent and longitudinal DIS cross sections (with unpolarized initial states) have been performed in Refs.~\cite{we4} and \cite{we7}, correspondingly. As was shown in \cite{we7}, large soft-gluon contributions to the cross sections $\sigma_L$ and $\sigma_T$ cancel each other in their ratio $\frac{\sigma_L}{\sigma_T}(x,Q^2)$ with an accuracy less than few percent at moderate $Q^2\lesssim m^2$ in the energy range $x>10^{-3}$. For this reason, soft-gluon resummation for the DIS cross sections with linearly polarized initial gluons within the TMD factorization scheme seems to be especially important. Possible cancellation of soft-gluon contributions of both unpolarized and linearly polarized gluons in the ratio ${\rm d}^3\sigma_L/{\rm d}^3\sigma_T$ would provide us an ideal, perturbatively stable observable for disentanglement of $f_1^{g}$ and $h_1^{\perp g}$. 

Note also that the quantity $R$ is sensitive to resummation of the mass logarithms of the type $\alpha_{s}\!\ln\left(Q^{2}/m^{2}\right)$ \cite{ACOT,Collins}. For this reason, it seems to be good probe of the perturbative heavy-quark densities in the proton \cite{we8,we8e}.

Concerning the experimental aspects, the quantity $R$ in charm and bottom production can be measured at the proposed EIC \cite{EIC} and LHeC \cite{LHeC2} colliders. Presently, the Callan-Gross ratio in heavy-quark electroproduction is practically unmeasured. Disentanglement of $\sigma_T$ and $\sigma_L$ (and, correspondingly, determination of the ratio $R$) requires a precise measurement of the $y$ dependence of the DIS  cross section, $\sigma\propto (\sigma_T+\varepsilon \sigma_L)$, where $\varepsilon=\frac{2(1-y)}{1+(1-y)^2}$. Unfortunately, both ZEUS and H1 experiments at DESY have not been able to provide precise results on $\sigma_L$. The reason is that data collected at HERA have small values of $y$. 

However, measurement of $F_L$ is included in the physics case for proposed LHeC \cite{LHeC2}. 
Designed LHeC luminosity and kinematic range will exceed the HERA values by a factor of thousand and twenty, respectively. Moreover, some principal improvements of the detector are planned which will allow to reach highest values of $y$ at much reduced background. (For more details, see review \cite{Klein}.) For this reason, one can hope that the LHeC (and EIC) will lead to the first precision measurement of $F_L$ and $R$.\footnote{As noted by Max Klein in Ref.~\cite{Klein}, "Based on the invaluable experience gained with H1 at HERA and on the design prospects for the LHeC and its $ep$ experiment, one can indeed be optimistic that Guido Altarelli's wish for a precise determination of $F_L$ will eventually be fulfilled."} 

\section*{Acknowledgements} The authors are grateful to S.~J.~Brodsky, A.~Kotzinian and   S.~O.~Moch for useful discussions. This work is supported in part by the ANSEF grant PS-nuclth-5027. 


\vspace{3mm}



\end{document}